# Predicting disease-related genes by path-based similarity and community structure in protein-protein interaction network

Ke Hu[1], Jing-Bo Hu[1], Ju Xiang[2,*], Hui-Jia Li[3-4,*], Yan Zhang[5,*], Shi Chen[5], Chen-He Yi[7]

[1]Department of Physics, Xiangtan University Xiangtan, Xiangtan 411105, Hunan China

[2]Neuroscience Research Center & Department of Basic Medical Sciences, Changsha Medical University, Changsha 410219, Hunan, China

[3]School of Management Science and Engineering, Central University of Finance and Economics, Beijing 100080, China

[4]Academy of Mathematics and Systems Science, Chinese Academy of Sciences, Beijing 100190, China

[5]Department of Computer, Changsha Medical University, Changsha 410219, Hunan, China

[6]School of Public Administration, Xiangtan University, Xiangtan 411105, Hunan, China

* Corresponding authors: Ju Xiang or Hui-Jia Li or Yan Zhang.

E-mail: xiang.ju@foxmail.com (J.X.); xiangju@aliyun.com (J.X.); hjli@amss.ac.cn (H.J.L.); zhangyancsmu@foxmail.com (Y.Z.); huke1998@aliyun.com (K.H); chenshi198001@qq.com (S.C.)

**Abstract:** Network-based computational approaches to predict unknown genes associated with certain diseases are of considerable significance for uncovering the molecular basis of human diseases. In this paper, we proposed a kind of new disease-gene-prediction methods by combining the path-based similarity with the community structure in the human protein-protein interaction network. Firstly, we introduced a set of path-based similarity indices, a novel community-based similarity index, and a new similarity combining the path-based similarity index. Then we assessed the statistical significance of the measures in distinguishing the disease genes from non-disease genes, to confirm their availability in predicting disease genes. Finally, we applied these measures to the disease-gene prediction of single disease-gene family, and analyzed the performance of these measures in disease-gene prediction, especially the effect of the community structure on the prediction performance in detail. The results indicated that genes associated with the same or similar diseases commonly reside in the same community of the protein-protein interaction network, and the community structure is greatly helpful for the disease-gene prediction.

**CONTENTS**

**1. Introduction**

    Identification of the hereditary disease-genes from human genome is one of the most important tasks in bioinformatics research[1, 2]. The traditional methods such as the positional cloning via linkage analysis were applied to discover the disease-related genes[3, 4], but they encounter difficulties, such as the variable disease penetrance [5] and the large number of genes among large family datasets that need to be analyzed[6, 7]. The former is an inherent problem that directly limits the performance of these traditional methods, while the latter is a labor-intensive task, which will cost much man power and resources. Therefore, it is necessary to develop the efficient computational approaches to predict unknown genes associated with certain diseases before the experimental research[8, 9].

    Up to now, many features and resources have been employed to predict disease-related genes, such as sequence features[10], gene functional annotations[11, 12], and protein-protein interaction (PPI) network[13, 14], etc. Among them, the coverage of the gene functional annotations is limited, though they play an important role in disease candidate gene prioritization. In the past decades, benefit from the rapid development of experimental technologies (e.g., yeast two-hybrid technology[15]), more and more PPI data are becoming available, which greatly promoted the study of the disease-gene discovery[16]. Various candidate gene prioritization methods have been proposed based on the PPI network analysis[12, 13, 17-20]. The theoretical basis of these network-based approaches is that genes associated with the same or similar disease phenotype, are not distributed randomly in the network, and they possess many common topological features, for example, they usually have high connectivity, reside in more central location of the PPI network and cluster together[21-23].

    Based on these topological properties, many scoring systems that match automatically these

topological features have been developed[24-29], for example, Xu and Li developed a classifier to predict disease genes by combining five different topological features[29]. A long-held and partially proved assumption shared by biologists is that genes associated with the same or similar diseases phenotypes are likely to be functionally related, and thus they commonly reside in the same neighborhood of molecular networks. For example, method based on direct neighbors of disease genes was very simple, but not the most efficient in disease-gene prediction[30]; then, Hsu et al considered the indirect neighbors on the basis of the direct neighbors of disease genes[31]; while Zhu et al proposed a kind of method integrating the direct neighbors, the indirect neighbors and the shortest path length between candidate and known disease genes[32]. It is worth noting that module structure is one of important properties in the PPI networks[33], while, in the previous studies, little attention was focused on the module property of the disease genes. It was clear that, proteins play their functions in a modular fashion, and mutations of proteins in the same module may lead to similar disease phenotypes[34]. The modular feature in the PPI network can be properly characterized by the community structure in network theory[35, 36]. Therefore, the community structure should be a more direct and robust property to capture a functional modularity in the PPI network. Further, it is expected that genes associated with the same or similar disease phenotypes commonly reside in the same community of the PPI network, and the community structure might be greatly helpful for disease-gene prediction.

In this paper, based on the PPI network, we firstly introduce a set of path-based similarity indices and a novel community-based similarity index, and then a new similarity combining the path-based similarity with community-based similarity is designed. Secondly, we assess the statistical significance of the measures in distinguishing the disease genes from non-disease genes, to confirm their availability in predicting disease genes. Finally, we apply these measures to the disease-gene prediction of single disease-gene family, and analyze the performance of these measures in disease-gene prediction, especially the effect of the community structure on the prediction performance in detail.

## 2. Datasets

### 2.1. Human PPI Datasets

Protein-protein interaction databases have become a major resource for investigating biological networks and pathways in cells. A number of publicly repositories for human PPI are currently available. Each of these databases has their own unique features with a large variation in the type and depth of their annotations. In this paper, human PPI data is taken from the Human Protein Reference Database (HPRD)[37]. All the information in HPRD has been manually extracted from the literatures by expert biologists who read, interpret and analyze the published data. We have downloaded a newest version (version number: Release 9) of the PPI data, in which the total number of genes annotated with at least one interaction is 9465 and the number of binary non-redundant human PPI is 37039. By analyzing the human PPI network, we revealed that the PPI network is composed of 110 connected components. The largest component contains 9219 proteins and 36900 interactions, while the other 246 proteins are distributed into 109 small components with 2-5 proteins. Since the majority of the small sub-networks contain only a few genes, it might not be of interest to check the distribution of the disease genes.

### 2.2. Disease-Gene Data

In order to evaluate the performance of our methods, a certain amount of known disease genes are required. In present paper, we employ HepatoCellular Carcinoma (HCC) genes that are extracted from OncoDB.HCC[38]. This data contains 605 significant genes that are concluded to be related to hepatocellular carcinoma by three criteria:

(1) Based on up-/down-regulated expression, genes were selected with at least three independent HCC microarray reports;

(2) Based on at least 2-fold expression changes within 70% patients, genes were selected with re-analyzed Stanford HCC microarray data;

(3) Genes were selected with web-lab experiments data.

Among the 605 disease genes, only 448 genes is distributed in the largest component of the present PPI network, and thus will be used for further analysis.

Both the basic topological features of the PPI network and the statistical properties of disease genes [39-41] in the main component of PPI network are summarized in Table 1. One can find that the disease-genes have higher connectivity, higher network efficiency and shorter distance than those of the non-disease genes. In addition, according to the Pearson correlation coefficient of degrees, disease genes exhibit an assortative mixing pattern of degrees, while the non-disease genes are disassortative mixing. This indicates that the disease genes tend to connect to other disease genes with similar degrees, while the non-disease genes with high degrees tend to connect to other non-disease genes with low degrees. From the difference of the degree heterogeneity, one can find that the degrees of the non-disease genes are more dispersive than those of the disease genes.

**Table 1:** Basic statistical properties of the main component of the PPI network and the HCC genes subnet. $N$ is the total numbers of nodes. $<k>$ is the average degree. $e$ and $<d>$ denote the efficiency and the average shortest distance between node pairs, respectively. $<C>$ and $r$ are the clustering coefficient and correlation coefficient of degrees, respectively. $H$ is the degree heterogeneity, defined as $<k^2>/<k>^2$.

| Measures | $N$ | $<k>$ | $e$ | $<d>$ | $<C>$ | $r$ | $H$ |
|---|---|---|---|---|---|---|---|
| Proteins (or genes) | 9219 | 8.005 | 0.252 | 8.419 | 0.342 | -0.036 | 4.411 |
| Disease genes | 448 | 18.085 | 0.271 | 7.310 | 0.195 | 0.080 | 3.770 |
| Non-disease genes | 8771 | 7.490 | 0.251 | 8.471 | 0.349 | -0.053 | 4.173 |

## 3. Methods

### 3.1. Definition of the topological similarity

#### 3.1.1. Path-based Similarity (PS)

The similarity between proteins (or genes) can be defined by using their essential attributes: two proteins are considered to be similar if they have the same or similar functions. However, available functional attributes are limited, and thus we focus on another type of similarity based on the PPI network topology. According to the used topological information, the topological similarity can be classified into two types: local similarity and global one[42-45]. Here, a set of path-based similarity indices, from local to global, are introduced.

(1) Path-based Similarity of 1st order (**PS1**): To measure the topological similarity between two genes in the PPI network, many approaches have been developed. Among them, the simplest approach may be the path-based similarity of 1st order, which is to assess whether two genes are connected directly in the network. If the two genes are directly connected by an edge, the similarity between them is defined as a constant of 1, otherwise zero. So, PS1 can be written as,

$$S_{xy}^{PS1} = A_{xy},$$ (1)

where $A_{xy}$ is the matrix element of the adjacency matrix $A$: $A_{xy}=1$ if nodes $x$ and $y$ are directly connected and $A_{xy}=0$ otherwise.

(2) Path-based Similarity of $2^{nd}$ order (**PS2**): In common sense, nodes $x$ and $y$ are more likely to have similar properties if they have many common neighbors. Hence, by combining PS1 with the common-neighbor feature, we can define the path-based similarity of $2^{nd}$ order as:

$$S_{xy}^{PS2} = A_{xy} + \beta(A^2)_{xy},$$  (2)

where $(A^2)_{xy}$ denotes the number of the common neighbors between nodes $x$ and $y$, which is also the number of different paths with length 2. In addition, $\beta$ is a free parameter of controlling the weights of paths with different lengths.

(3) Path-based Similarity of $3^{rd}$ order (**PS3**): Above, PS2 has low computational complexity while they may still miss some useful information for underlying topological similarity, because they consider only the local paths with length 2. Therefore, we further consider paths with length 3, and define the path-based similarity of $3^{rd}$ order:

$$S_{xy}^{PS3} = A_{xy} + \beta(A^2)_{xy} + \beta^2(A^3)_{xy},$$  (3)

where $(A^3)_{xy}$ denotes the number of different paths with length 3 between nodes $x$ and $y$. In principle, one can further define the path-based similarity of higher order by considering the longer paths, while it may not be always necessary in general networks due to the small-word effect[39].

(4) Global Path-based Similarity (**PSG**): For comparison, we finally consider the *global* similarity which is based on all the paths between two nodes in the network, which can be written as:

$$S_{xy}^{PSG} = \sum_{n=1}^{\infty} \beta^{n-1}(A^n)_{xy},$$  (4)

where $(A^n)_{xy}$ is equal to the number of paths of length $n$ from node $x$ to node $y$.

In fact, PSG is equivalent to the Katz index that has been widely applied to characterize node similarity in complex networks[46]. In order to guarantee the convergence of PSG, the parameter $\beta$ must be smaller than the reciprocal of the maximum eigenvalue of the adjacency matrix $A$ [43]. In this paper, we fixed $\beta = 0.01$ for the three path-based similarities, PS2, PS3, and PSG. Of course, one can tune $\beta$ to find optimal value corresponding to the highest prediction accuracy, however this optimal value is different for different similarity indices and one parameter-dependent measure is less practical in dealing with huge-size network since the tuning process may take much time.

### 3.1.2. Community-based Similarity (CS)

The PPI network shares many common properties with many other real-world networks, such as the small-world property, scale-free feature, and community structure [39]. Among them, community structure has been recognized as an important bridge to connect the topological structures and functional modules. It has been widely studied in various real-world networks such as the Internet, the word wild web, epidemiology, metabolism, ecosystems[33, 35, 39, 47, 48]. Here, we introduce the community structure into the disease-gene prediction problem, aiming to further improve the performance of the network-based disease-gene prediction.

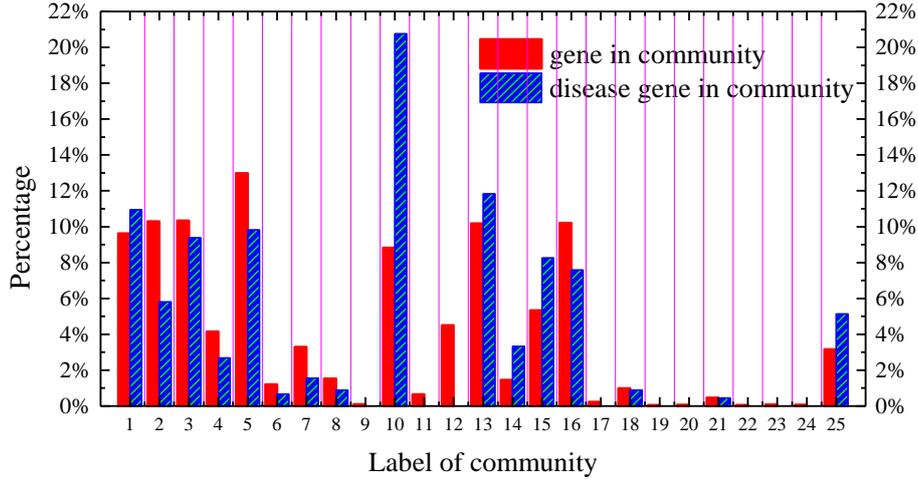

**Figure 1:** The percentages of genes and disease genes in each community.

Many methods have been recently developed to detect community structures in complex networks[49-51]. Here, the community structures in the PPI network are extracted by the BGLL method[52], because it has low computational complexity and high accuracy in community detection. The present PPI network is naturally divided into 25 communities and the corresponding modularity, which quantifies the modular property of the network, is equal to 0.533. And then we aggregate the genes and disease genes in each community and calculate their percentages with respect to the total candidate genes and the total disease-genes, respectively. The statistical results in these 25 communities are depicted in figure 1. One can find that the HCC disease-related genes should be functionally correlated and do not distributed randomly among the communities of the PPI network. For instance, about 77.7% of HCC disease-related genes are mainly distributed within 7 communities, and there are 8 communities without HCC disease-related gene. And more than 20% of disease genes are clustered in the $10^{th}$ community which has only 9% of genes in the whole network, while the $10^{th}$ community should include only 4.86% of the disease genes, if the disease genes are randomly distributed among the communities.

The non-random distribution of disease-genes in communities might be very helpful for the disease-genes prediction. Institutively, in real-world networks, two individuals within the same community should be more similar than in different communities. For example, in the interpersonal relationship networks, community formation may be based on the human career, age and other factors; in the metabolic network and neural network, communities may correspond to the functional units. In order to extract the useful information in the community structure, we defined the community-based similarity as:

$$S_{xy}^{CS} = \begin{cases} S_0, & C_x = C_y; \\ 0, & C_x \neq C_y, \end{cases} \tag{5}$$

where $C_x$ is the label of the community to which node $x$ belongs. We define $S_0 = 1/N_x$, where $N_x$ is the total number of genes in the community to which node $x$ belongs. This means that the smaller the size of the community is, the more similar the genes in the community are. The reason of this consideration will be discussion in the following section about similarity score.

### 3.1.3. Combined similarity based on path structure and community structure

Based on path structures and community structure, a combined similarity of PS and CS (denoted by PS-CS) can be defined as:

$$S_{xy}^{PS-CS} = (1-\alpha)S_{xy}^{PS} + \alpha S_{xy}^{CS}, \qquad (6)$$

where $\alpha \in [0,1]$ is a free parameter to adjust the contribution of the paths and community structure to the similarity. If $\alpha = 0$, the community does not contribute to the similarity, i.e., the similarity indices will reduce to original PS. If $\alpha = 1$, it will be reduced to the CS.

### 3.2. Similarity scores of genes with respect to the disease genes

To prioritize the candidate disease genes, the similarity *Score* of a candidate gene $x$ with respect to the disease gene set is calculated by summing the similarity of the candidate gene $x$ with all known disease genes. For the path-based similarities, the similarity scores of candidate genes are written as:

$$Score(x, PS) = \sum_{y \in D} S_{xy}^{PS}, \qquad (7)$$

where $D$ denotes the disease genes set. For the community-based similarity, the score can be expressed as:

$$Score(x, CS) = \sum_{y \in D} S_{xy}^{CS} = \frac{n_x}{N_x}, \qquad (8)$$

where $n_x$ denotes the number of disease genes in the community to which gene $x$ belongs. In fact, $n_x/N_x$ is the abundance of the disease genes in the community. We employ it to reflect the degree of correlation between the community and the disease phenotype, i.e., if the abundance of the disease genes in a community is relatively high, this community may be considered to be closely related to the disease and the candidate genes in it will be more possibly related to the disease.

According to the combined similarity of PS and CS, the similarity *Score* is then written as

$$\begin{aligned} Score(x, PS-CS) &= \sum_{y \in D} S_{xy}^{PS-CS} \\ &= (1-\alpha)\sum_{y \in D} S_{xy}^{PS} + \alpha \sum_{y \in D} S_{xy}^{CS} \\ &= (1-\alpha) \cdot Score(x, PS) + \alpha \cdot Score(x, CS) \end{aligned} \qquad (9)$$

Finally, the candidate genes are sorted by the scores in decreasing order. The top predicted genes are more likely to be disease-related genes.

### 3.3. Metric

To estimate the performance of the prediction algorithms, the general process is that the known disease genes are randomly divided into several subsets. In each independent realization, one of these subsets is treated as test set, and remain subsets are put together to build a training set. And then we calculated the similarity score of every gene with respect to the genes in the training set. According to the scores, the genes will be sorted in decreasing order: The top genes are more likely related to the disease.

We adopt two standard measures to determine the accuracy of disease-gene prediction methods: AUC (area under the receiver-operating characteristic curve) and precision. The AUC evaluates the performance based on the receiver-operating characteristic (ROC) curve which can intuitively reflect the full-scale performance according the whole score list[53]. It is a two-dimensional graph in which true-positive rate is plotted on the *Y* axis and false-positive rate is plotted on the *X* axis. The true-positive rate (also called sensitivity or recall) and the false-positive rate (i.e., 1-specificity) is defined as

$$Ture\ positive\ rate = Sensitivity = \frac{TP}{TP+FN}, \qquad (10)$$

and,

$$\textit{False positive rate} = 1 - \textit{Specificity} = \frac{FP}{FP + TN}, \tag{11}$$

where TP, FP, TN and FN represent true positive, false positive, true negative and false negative, respectively. By calculating the area under an ROC curve (i.e., AUC), one can obtain a single scalar value representing the comprehensive performance. Because AUC is a portion of the area of the unit square, its value will always be between 0 and 1. If AUC is smaller than 0.5, the prediction method will be completely ineffective. In a statistical sense, AUC could be considered as the probability that the score of a randomly chosen positive sample is higher than that of a randomly chosen negative sample.

Different from AUC, precision only focuses on the top-$L$ genes with highest scores. It is defined as the ratio of relevant items in the top-$L$ list, i.e., the true-positive rate among selected top-$L$ genes. That is to say, among the top-$L$ genes, if $L_r$ genes are accurately predicted, the precision is calculated by

$$\text{Pr}\,ecision = \frac{L_r}{L}. \tag{12}$$

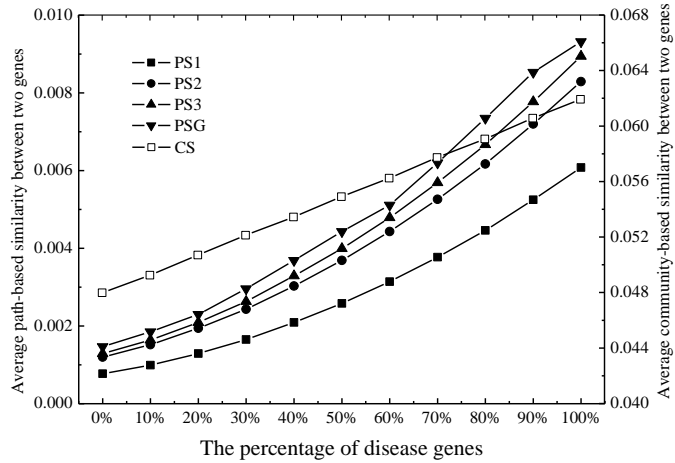

**Figure 2**: The average topological similarity between genes by PS1, PS2, PS3, PSG and CS as a function of the percentage of disease genes in the substituted set. In order to avoid sampling bias, the random sampling is repeated 1000 times for each percentage of disease genes. Each point corresponds to the average of the 1000 samples.

## 4. Experimental results

### 4.1. Analysis of feasibility

In order to demonstrate the feasibility of our methods, we perform two different statistical analyses. On the one hand, we are to test whether the average level of the topological similarity (PS and CS) between disease genes is clearly different from that between non-disease genes as well as that between non-disease gene and disease gene. The average topological similarity between genes in the disease-gene set is calculated by

$$\overline{S} = \frac{\sum\limits_{x, y \in D, y \neq x} S_{xy}}{n_D(1 - n_D)}, \tag{13}$$

where $S_{xy}$ is referred to the similarity about PS1, PS2, PS3, PSG or CS. $D$ and $n_D$ denote the disease-gene set and the number of disease genes in the disease-gene set, respectively. And then, we randomly select a certain percentage of disease genes from the disease-gene set $D$ and replace them by the non-disease genes selected randomly from the non-disease gene set. This will generate a substituted disease-gene set that contains remaining disease genes and selected non-disease genes. According to equation (13), we calculate the average similarity between genes in the substituted disease-gene set with different percentages of remaining disease genes, which is depicted in figure 2. In figure 2, it is clearly showed that the more the disease genes in the substituted disease-gene set, the higher the average similarity is. For example, when the percentage of the remaining disease genes is 100%, i.e., disease genes are not replaced by non-disease genes, the average similarities is significantly higher than those of all disease genes replaced by non-disease genes (i.e., the case of 0%). This indicates that the defined similarities above (PS and CS) should be able to effectively distinguish the disease genes from the candidate genes.

On the other hand, the average similarity scores for all known disease genes and all candidate genes have been computed by using the path-based similarity (PS) and the community-based similarity (CS) (Table 2). Based on the definition of the similarity score, the average scores with respect to the disease gene set are calculated by

$$\overline{Score} = \begin{cases} \dfrac{\sum_{y \in D} Score(y)}{n_D} & \text{for disease genes} \\ \dfrac{\sum_{y \in \eta} Score(y)}{n_\eta} & \text{for candidate genes} \end{cases}, \tag{14}$$

Where $\eta$ and $n_\eta$ denote the candidate gene set and the number of genes in the set, respectively. Table 2 shows that, for each topological similarity, the average similarity score of disease genes is obviously higher than that of candidate genes. Therefore, one could expect that the prediction methods based on these similarities can help us to identify the disease genes from candidate genes.

**Table 2:** The average similarity scores of disease genes and candidate genes with respect to the disease gene set, for different similarity definitions.

| Similarity | Average Score (disease genes) | Average Score (candidate genes) |
|:---:|:---:|:---:|
| PS1 | $6.08 \times 10^{-3}$ | $0.77 \times 10^{-3}$ |
| PS2 | $8.29 \times 10^{-3}$ | $1.20 \times 10^{-3}$ |
| PS3 | $8.94 \times 10^{-3}$ | $1.29 \times 10^{-3}$ |
| PSG | $9.32 \times 10^{-3}$ | $1.47 \times 10^{-3}$ |
| CS | $6.19 \times 10^{-2}$ | $4.80 \times 10^{-2}$ |

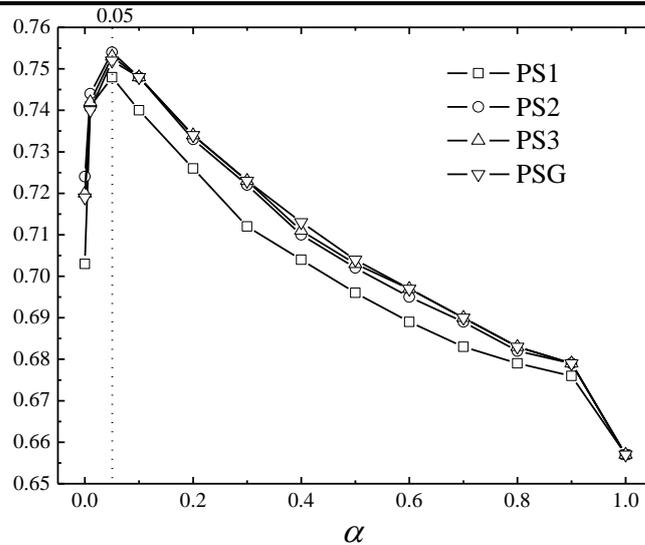

**Figure 3:** The AUC values as a function of $\alpha$ for each PS-CS.

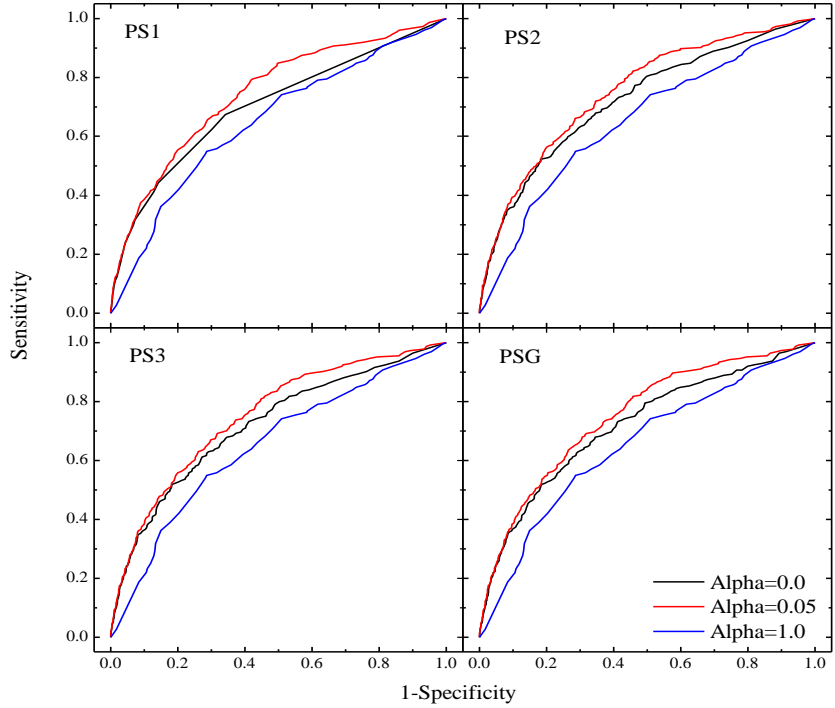

**Figure 4:** ROC curves from cross validations, by using different PS-CS with different $\alpha$ values, to predict HCC disease genes in the human PPI network.

## 4.2. Performance of method

### 4.2.1 ROC and AUC

From the above statistical analysis, one can confirm that both the PS and CS are effective for predicting the disease genes. Naturally, it should be considered whether the combined similarity can give a better performance. Further, whether there exists the optimal the parameter $\alpha$ controlling the weights of PS and CS. In order to answer these questions, we calculate the AUC for different $\alpha$ (see figure 3). Interestingly, the prediction accuracy measured by AUC shows a maximum when $\alpha \approx 0.05$. Moreover, the optimal prediction accuracy is significantly higher than both those of single PS and CS. This indicates that better results are obtained by the combination of these similarities. In the following analysis, we focus on the three cases of $\alpha = 0.0$, $\alpha = 0.05$ and $\alpha = 1.0$, corresponding to PS, the optimal combination of PS-CS as well as CS. The corresponding ROC curves and AUC values are showed in figure 4 and table 3.

### 4.2.2. Precision

To estimate the Precision of the methods, 50% of disease genes (224 genes) are randomly selected as training set, from the HCC disease-gene set. The remaining 224 disease genes mixing with a part of original non-disease genes are used as the probing set. Because the too large or too small sample sizes are not appropriate for statistical analysis, we randomly select 3076 non-disease genes in the PPI network. The prediction methods predict the disease genes from these 3300 (224+3076) candidate genes. The top L=165 genes (i.e., 5% of the candidate genes) are predicted as the disease genes. In order to avoid sampling bias, the random sampling above is repeated 1000 times and the results are showed in table 4.

Similar to AUC, Table 4 shows that, by considering the community structure, the precision of the path-based similarity can be improved markedly. Without the community structure, the highest precision is obtained by PS2, and the highest precision is also obtained by PS2-CS, after considering the community structure. Moreover, the maximum improvement of precision appears in PS3, from 0.263 to 0.289, increasing by 10.0%, though AUC of PS3 (Table 2) is only increased by 4.6%. It means that after the addition of the community structure, the overall performance improvement (AUC of PS3) is not very obvious, but the quality of disease-gene prediction trends to move forward. Finally, in any case, it is clear that the community structure can be able to be helpful for the prediction of disease genes in the human PPI networks.

Table 3: Prediction accuracies measured by AUC.

| Similarity | AUC | | |
|------------|----------------|----------------|----------------|
|            | $\alpha = 0.00$ | $\alpha = 0.05$ | $\alpha = 1.00$ |
| PS1-CS     | 0.703          | 0.748          | 0.657          |
| PS2-CS     | 0.724          | 0.754          | 0.657          |
| PS3-CS     | 0.720          | 0.753          | 0.657          |
| PSG-CS     | 0.719          | 0.752          | 0.657          |

Table 4: Prediction accuracies measured by Precision.

| Similarity | Precision | | |
|------------|----------------|----------------|----------------|
|            | $\alpha = 0.00$ | $\alpha = 0.05$ | $\alpha = 1.00$ |
| PS1-CS     | 0.258          | 0.280          | 0.178          |
| PS2-CS     | 0.266          | 0.291          | 0.178          |
| PS3-CS     | 0.263          | 0.289          | 0.178          |
| PSG-CS     | 0.262          | 0.286          | 0.178          |

### 4.2.3. Discussion

From Table 3 and 4, we can find that the performance of prediction algorithm based on PS is significantly better than that for CS. According to the definition of CS, the similarity between genes within the same community is the same, which will cause that a large number of gene pairs are assigned identical similarity. If we regard the similarity as the energy that is assigned to the gene pairs, then many gene pairs crowed into very few energy levels. Taking the present detected communities as an example, there are at most 25 energy levels, since the PPI network are divided into 25 communities. Therefore, the performance of the CS method is possibly rooted in the high degeneracy of the energy status. However, the PS could break the degeneracy, and give the relative good performances. Among the four PS methods, the highest AUC value has been obtained by PS2, and the lowest one is given by PS1. Although PS1 requires only information on the nearest neighbors and thus has very low computational complexity, while the information generally seems to be insufficient. On the other hand, the PSG has the highest computational complexity, but its prediction accuracy is not beyond those of PS2 and PS3. Perhaps, because the PPI network shows strong small-world effect, considering the 2-order or 3-order local paths for the disease-gene prediction seems to be enough. This situation is very similar to the link prediction based on the local path similarity index, where it is also confirmed that the global path-based similarity is not always necessary for some real networks with strong small-world effect[54].

When comparing the AUC and Precision of the combined similarity ($\alpha = 0.05$) with that of PS

( $\alpha = 0.0$ ) and CS ( $\alpha = 1.0$ ), a great improvement can be obtained due to the combination of PS and CS. The excellent performance of the methods may result from that the non-random distribution of disease genes in communities can provide very useful information for the prediction of the disease-gene family. On the one hand, the genes in the same community often have higher similarity than between communities, and especially, the similarity scores of the candidate genes with respect to the disease gene set, by the definition of CS, exactly corresponds to the abundance of disease genes in the communities. Generally, if the known disease genes related to the certain disease are very scarce or even absent in a community, the possibility of the candidate genes in this community being the disease genes may be very low; vice versa, if the relative number of disease genes is very large in a community, the candidate genes in this community are more possible to be related to the disease. Therefore the information of community structure could be helpful for the identification of the disease genes in the PPI network.

Theoretically, the combination of different similarities may be particularly relevant for disease gene prediction, if they are designed independently by different methods. In our case, the PS and CS were designed to assess the similarity of genes in PPI network from different while related sides. In fact, there exists correlation between PS and CS. For example, the similarity indices such as PS can be applied to community detection, where the node pairs with high similarity are generally clumped into communities[42]. However, the two similarities should not be correlated strongly too, otherwise they will generate the same or similar results in the disease-gene prediction. In our previous work[55], a network-based prediction method has been developed by using the global path-based topological similarity and confirmed that path-based similarity could give a relatively good performance for disease genes prediction. In present work, the CS has been introduced as a novel feature to assess modularity of proteins in the PPI network. The introduction of the CS further enhanced the prediction performance, which indicates that the community structure may reflect some features of disease genes in the PPI networks. The information from PS and CS could complement each other, and thus the combination can finally lead to better performance for predicting disease genes in the human PPI network.

## 5. Conclusion

In this paper, in order to predict disease genes in the human PPI network, a set of path-based similarity methods and community-based similarity methods were introduced, and then a kind of combined similarity methods was proposed by considering the path structure and community structure in the networks. We analyzed the feasibility of the similarity methods for predicting the disease genes in the PPI network, and investigated the effect of the community structure on the prediction of the disease genes. The results indicated that the performance of the prediction methods can be greatly improved due to the introduction of the community structure, though the alone community-based similarity method is not good at predicting the disease genes in the PPI networks.

In general, proteins seldom function independently, while rather in a modular fashion. When combining the path-based similarity with the community-based similarity, the prediction performance not only improves, but the advantage of the low computational complexity of the methods (such as PS1, PS2 and PS3) also preserves. From the above results, one could validate that the community structure is a significant feature for improving the performance of disease-gene prediction. The genes related to the disease may be indeed clustered into modules in the PPI network. Thus, exploring the modular feature of proteins in the PPI network would be considerably important for both developing effective disease-gene prediction methods and understanding the underlying functions of proteins.

Moreover, we also have noted that the disease genes may be situated at the border of communities, while not inside, which may also be helpful for the prediction of disease genes. In the future, the distribution of disease genes in communities should be deserved to further study in predicting disease genes. Finally, we believe that community structure, as a relevant and intuitional feature for characterization of networks, should be able to easily combine with other methods so as to predict disease genes more effectively.

## Acknowledgement

This work has been supported by the construct program of the key discipline in Hunan province, the Scientific Research Fund of Education Department of Hunan Province (Grant No. 17A024), the Scientific Research Project of Hunan Provincial Health and Family Planning Commission of China (Grant No. C2017013), the Scientific Research Fund of Education Department of Hunan Province (Grant No. 17C0180, 17B034, 15C0164 and 14B024), and the Department of Education of Hunan Province (Grant No. 15A023), the Hunan Provincial Natural Science Foundation of China (Grant No. 2015JJ6010), the National Natural Science Foundation of China (Grant No. 11404178 and 71401194), and the National Social Science Foundation of China (Grant No. 16BGL177).